# Charge Acceleration and Field-Lines Curvature: A Fundamental Symmetry and Consequent Asymmetries

Avshalom C. Elitzur[a], Eliahu Cohen[b*] and Paz Beniamini[c]

[a] *Iyar, The Israeli Institute for Advanced Research, Rehovot, Israel.*
[b] *School of Physics and Astronomy, Tel-Aviv University, Tel-Aviv 69978, Israel.*
[c] *Department of Physics, The Hebrew University, Jerusalem 91904, Israel.*
*eliahuco@post.tau.ac.il*

When a charge accelerates, its field-lines curve in a typical pattern. This pattern resembles the curvature induced on the field-lines by a neighboring charge. Not only does the latter case involve a similar curvature, it moreover results in attraction/repulsion. This suggests a hitherto unnoticed causal symmetry: charge acceleration ⇔ field curvature. We prove quantitatively that these two phenomena are essentially one and the same. The field stores some of the charge's mass, yet it is extended in space, hence when the charge accelerates, inertia makes the field lag behind. The resulting stress in the field stores some of the charge's kinetic energy in the form of potential energy. The electrostatic interaction is the approximate mirror image of this process: The potential energy stored within the field turns into the charge's kinetic energy. This partial symmetry offers novel insights into two debated issues in electromagnetism. The question whether a charge radiates in a gravitational field receives a new twist: If all the charge's field-lines end with opposite charges that also resist gravity, no radiation is expected. Similarly for the famous absence of a physical manifestation of the Maxwell equations' advanced solution: Just as Einstein argued, the reason for this absence is probabilistic rather than reflecting some inherent time-asymmetry. Despite the apparent equivalence between the "ontological" and "instrumentalist" viewpoints concerning the physical reality of field-lines, there may be cases in which their experimental predictions differ.

## 1. ELECTRIC FIELD-LINES: MATHEMATICAL ABSTRACTIONS OR PHYSICAL ENTITIES?

Ever since Faraday has introduced the concept of "field" and its concomitant "lines of force," the nature of these entities has remained elusive. Faraday himself is known to have held a strong "ontological" viewpoint, conceiving of the field-lines as real, thin "bands." Even the mathematically-minded Maxwell [1] has explicitly followed him on this issue. Most modern physicists, however, prefer the more prudent, "instrumentalist" viewpoint, regarding the field-lines as imaginary mathematical abstractions. With the advent of the Maxwellian account of the electromagnetic wave as a disturbance proceeding along the field-lines, interest in the field-lines themselves has dwindled, and the entire realm of electrostatics has been overshadowed by electromagnetism.

However, to quote Harpaz, "[F]ields are not only pedagogical or technical concepts that help to describe and calculate physical processes. Fields are independent physical entities, as Einstein has suggested [2], and they should be treated accordingly" [3, p. 222]. This observation invites a closer inspection of the field-lines as well.

Indeed, while textbooks often caution that field-lines "do not physically exist," they nevertheless ascribe them some highly visualized properties, *inter alia*: (*i*) Field-lines begin and end only at charges (by convention, beginning at + and ending at −) or at infinity. (*ii*) They never cross. (*iii*) They are closer together where the field is stronger. (*iv*) At any location, the direction of the electric field is tangent to the electric field-line that passes through that location. (*v*) They possess longitudinal tension ("pull their ends to become shorter"). (*vi*) They exert lateral pressure ("repel one another"). These properties are supposed to explain all the familiar-yet-intriguing electrostatic phenomena, like attraction and repulsion.

Field-lines, moreover, exhibit the same kinematic behavior as other physical bodies. When the charge moves inertially, the field-lines exhibit the same inertial motion (indeed, their inertial lag underlies the electromagnetic radiation, see below). Furthermore, upon nearing *c*, they undergo the familiar Lorentz transformation (Fig.1b):

$$\vec{E}'_\parallel = \vec{E}_\parallel \ , \ \vec{B}'_\parallel = \vec{B}_\parallel \ , \ \vec{E}'_\perp = \gamma(\vec{E}_\perp + \frac{\vec{v}}{c} \times \vec{B}) \ , \ \vec{B}'_\perp = \gamma(\vec{B}_\perp - \frac{\vec{v}}{c} \times \vec{E}) \qquad (1)$$

where $\parallel$ and $\perp$ denote parallel and perpendicular components to the velocity $\vec{v}$.

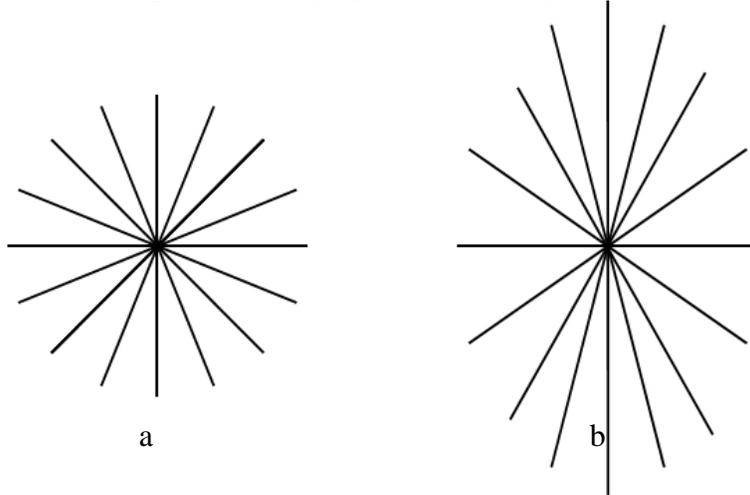

**FIGURE 1.** Field-lines of a charge (a) at rest, (b) moving with a constant relativistic velocity.

On the other hand, when field-lines move, magnetic field-lines emerge according to Maxwell's equations, perpendicular to both the electric line and its motion. This is a property unique to the electric field, for which a simplistic mechanical approach is not suitable.

In what follows we study the field-lines' dynamics from the "ontological" perspective, viewing them as physical objects endowed with mechanical properties like those listed above, while keeping in mind that other properties remain ill-understood. We then venture to derive some novel insights into the foundations of electromagnetism. It is also possible that the two approaches concerning the reality of field-lines eventually yield different experimental predictions.

## 2. FIELD-LINES CURVATURE: ELECTROMAGNETIC

When the charge's motion undergoes uniform acceleration, the lines assume a special form: They *curve* opposite to the motion. This pattern stems from the locality restriction: Time is required for every change in the charge's motion to be transferred to the field-lines' remote ends. Consider first a single velocity change: The old field-line (being, *e.g.*, at rest together with the charge) vanishes, from inside out, at velocity *c*, while the new field-line emerges, from inside out, at velocity *c* (but now [see Sec.1] being in inertial motion relatively to the previous line). Between the two lines' ends, a "kink" reconnecting them keeps widening (Fig.2a). The same process occurs in the perpendicular magnetic field. This is Maxwell's electromagnetic wave.

Next consider *ongoing* acceleration (2b): Assuming it is uniform, the momentary straight segments of the field-line become infinitesimally short, while the lags between them widen, hence the reconnecting "kinks" become longer until comprising the entire field-line, stretching to the direction opposite to the charge's motion (see [4] and [5] for detailed accounts).

The resulting electromagnetic field has two components: The "velocity field" depending on $1/R^2$, and the "acceleration field" depending on $1/R$ [6]. Therefore, over long distances, the latter becomes dominant.

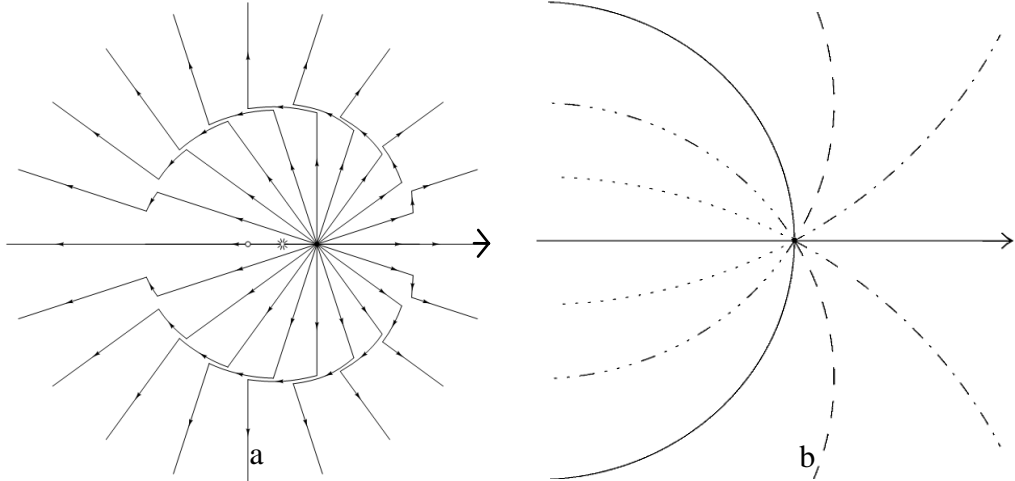

**FIGURE 2.** (a) Field-lines of a charge after a single velocity change: (b) Field-lines of a charge under constant acceleration.

Another way to understand this curvature is by deriving it from the famous $c$. Let us view the single field-line as a vibrating cord between a charge $q$ and an opposite charge $Q$, separated by $r$. The speed at which the disturbance propagates along this line is given by the well-known relation:

$$v = \sqrt{\frac{T}{\rho}} = \sqrt{\frac{F}{\varepsilon_p / rc^2}} = \sqrt{\frac{Qq/r^2}{Qq/r^2 c^2}} = c, \qquad (2)$$

where $T$ is the tension (Coulomb force in this case), $\rho$ the mass density per unit length, and $F$ and $\varepsilon_p$, respectively, are the system's electric force and potential energy.

As (2) is the velocity by which the field's is updated about the charge's velocity, in the charge's vicinity the field-lines assume the typical curvature [7]:

$$R_c \cong \frac{c^2}{a \sin \theta}, \qquad (3)$$

where $R_c$ is the curvature's radius, $a$ the acceleration and $\theta$ the angle between the acceleration's and the field-lines' directions.

An important distinction is in order here, not explicitly stated by other authors, who seem to take it for granted, yet overlooking it leads to confusions. In order for the above curvature to be of a purely kinematic origin, different from the electrostatic curvature to be described in the next Section, the charge must accelerate by *non-electrical means*. Otherwise, the additional electric field would obscure the picture. How, then, can a charge be accelerated without the aid of another electric field? If the charge is macroscopic, it may be pushed mechanically. Such a push is also basically electric, caused by the electron shells of the pushing bodies' atoms, but for a large charge these microscopic fields are negligible. With microscopic charges, there are cases like beta particle emission, where the electron is pushed mainly by non-electric forces. In short: field-lines curvature occurs when the charge alone is accelerated, making its field lag behind.

# 3. FIELD-LINES CURVATURE: ELECTROSTATIC

It is the above field-lines' curvature due to acceleration that motivates this paper's main thesis. For a strikingly similar curvature appears in a much commoner case, well-known ever since Faraday. When two opposite/like charges come close, they attract/repel one another by the Coulomb force

$$F = \frac{q_1 q_2}{r^2}, \qquad (4)$$

which, with the conventional use of test-particles, shows their field-lines curving, respectively, towards/away from each other (Fig.3).

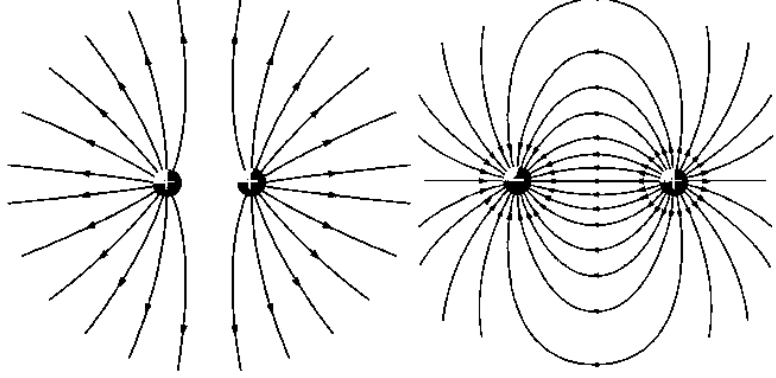

**FIGURE 3.** Electrostatic field-lines curvature.

Mathematically, this curvature follows the simple superposition principle: The net electric field produced by a system of charges is equal to the vector sum of all individual fields, produced by each charge at that point. For two charges, the total field is

$$\vec{E} = \vec{E}_1 + \vec{E}_2 = \frac{Q_1}{r_1^3} \cdot \vec{r}_1 + \frac{Q_2}{r_2^3} \cdot \vec{r}_2, \qquad (5)$$

which may, in fact, be derive from (4) as follows. Using a well-known formula from differential geometry for the curvature $\kappa$ of parametric curve, $\vec{\gamma}$ and its first derivatives:

$$\kappa = \frac{|\vec{\gamma}' \times \vec{\gamma}''|}{|\vec{\gamma}'|^3}, \qquad (6)$$

we derive the radius of curvature for the total field of two point charges

$$R_c(r) = \frac{|\vec{E}|^3}{\left|\vec{E} \times \left(\frac{\partial E(\vec{r})}{\partial \vec{r}}\right)\vec{E}\right|} =$$

$$\frac{\left|\frac{q_1(\hat{r}-\hat{R}_1)}{(r-R_1)^2} + \frac{q_2(\hat{r}-\hat{R}_2)}{(r-R_2)^2}\right|^3}{\left|\left[\frac{q_1(\hat{r}-\hat{R}_1)}{(r-R_1)^2} + \frac{q_2(\hat{r}-\hat{R}_2)}{(r-R_2)^2}\right] \times \left\{\left(\frac{\partial E(\vec{r})}{\partial \vec{r}}\right)\left[\frac{q_1(\hat{r}-\hat{R}_1)}{(r-R_1)^2} + \frac{q_2(\hat{r}-\hat{R}_2)}{(r-R_2)^2}\right]\right\}\right|}, \qquad (7)$$

where $R_1$ and $R_2$ are the two charges' positions and $\frac{\partial E(\vec{r})}{\partial \vec{r}}$ the matrix of the two charges' field derived by $\vec{r}$.

In short, when field-lines' curvature is produced electrostatically, *acceleration* follows, to the opposite side of the acceleration which in Sec.2 gave rise to the electromagnetic curvature.

# 4. THE SYMMETRY CONJECTURE: CHARGE ACCELERATION ⇔ FIELD-LINES CURVATURE

A new causal symmetry, very simple yet equally profound, thereby suggests itself: *electromagnetic curvature (the charge pulls the field) = electrostatic curvature (the field pulls the charge)* (8)

Is this symmetry genuine? If we follow Harpaz [3] and Rowland [8] in viewing the field as an independent physical entity, then, within the single charge-plus-field system, relation (8) manifests time-reversal symmetry [9], inherent to all physical interactions. In other words, "A charge accelerates, whereby its field follows it" entails "A field accelerates, whereby its charge follows it." As can be seen in [6], these two processes involve, radiation emission. In terms of Newton's third law, we may even refine this reciprocity to the interaction between the charge $c$ and its field-line $l$:

$$\sum F_{c,(l_1,l_2...l_n)} = -\sum F_{(l_1,l_2...l_n),c} . \tag{9}$$

The only question, of course, is: How can a field be accelerated first? to which our answer would be the Symmetry Conjecture itself: The electrostatic interaction is the case where the fields act on each other first, their charges following them.

Moreover, bearing in mind the enormous spatial difference between these two entities – the elementary charge being microscopic or even point-like while the field extends over space and interacts with numerous other bodies – *it is only natural for this symmetry to be weak, hardly noticeable in other than ideal situations*.

Further support for our conjecture comes from the fact that the electric field, among its "mechanical" properties mentioned above, is endowed with *mass and stress*, again rendering it a causal agent in itself. Following is a discussion of these properties that will later facilitate a rigorous proof for the Symmetry Conjecture.

## 4.1. The Field's Mass

That the electric field has mass is a relativistic consequence of its possessing energy:

$$m_{field} = \frac{1}{c^2} \frac{1}{8\pi} 4\pi \int_{r_q}^{\infty} \frac{q^2}{r^2} dr = -\frac{q^2}{2c^2} \frac{1}{r}\bigg|_{r_q}^{\infty} = \frac{q^2}{2c^2 r_q} \tag{10}$$

which will appear later in the definition of the classical charge radius. Harpaz and Soker [10] further ascribe the electric field *inertia*, a natural consequence of its mass. Moreover, assuming no action-at-a-distance, the field must act as a carrier of momentum between the charges. This role of the field provides a natural explanation for several cases where both electric and magnetic forces operate together in ways that seem to defy Newton's third law [11].

But then, if the accelerating charge imparts momentum to its field, and if the field has inertia, then, by momentum conservation, the field must exert an equal and opposing force on its charge. We would therefore expect the charge's acceleration to be reciprocated by *deceleration*. In reality, however, this deceleration is hardly noticeable, the reason being simple but important: *Most of the charge's mass resides within the charge itself, while only little is spread over the field, even less in its farthest parts*. Indeed, while Eq. (9) subjects the charge-field interaction to Newton's Third Law, it will be the application of the Second Law, $F=ma$, that will later illuminate the electrostatic curvature – see Eq.(15) below.

Based on this unequal distribution of the charge's mass, Rowland [8] likens the accelerating charge to a rod which is mechanically pushed at its middle: The initial force can be relatively weak because the push's effect has not yet reached the rod's ends. Only as the push's effect propagates along the rod, it becomes harder to keep accelerating it. Similarly for the charge: When only the charge itself is pushed, it takes time for the push to affect the surrounding field, hence the field's inertial resistance joins the particle's resistance only gradually. An earlier account of this dynamics was given by Feynman [12].

In summary, it is the field's inertia that makes it lag behind the accelerating charge, giving rise to the field-lines' curvature to the opposite direction.

## 4.2. The Field's Stress Force

As early as Maxwell's original work [1] the electromagnetic forces were pictured as mediated by a medium which is subject to stress. This also follows naturally from the "lateral pressure" by which the field-lines appear to repel one another (Sec.1). The field's stress is often employed as a pedagogical aid for teaching electromagnetism [13] [14], but also as a research tool for understanding the field concept itself [3] [8].

"Stress," denoting the internal forces per unit area, can be described by the second-order tensor $\sigma_{ij}$. It constitutes the spatial components of the stress-energy-momentum tensor: $T_{\mu\nu}$. "Strain" measures the deformation, i.e., relative change in shape or size of an object, due to an external force. In the general case it is also a tensor quantity.

Applied to our case, field-lines curvature is the deformation that the electric field undergoes due to external force. Indeed, the fact that the field-lines are spread so as to maintain maximal mutual distances indicates that stress is present even when no external force is applied. Anything that makes the lines curve, therefore, creates additional stress.

Harpaz and Soker [10] define the stress force of an electric field:

$$F_s = \frac{E^2}{4\pi R_c} = \frac{E^2 a \sin\theta}{4\pi c^2} \tag{11}$$

for which "force density" would perhaps be a more appropriate term, and which, we suggest, is strictly connected to the electromagnetic stress tensor as presented by Hermann [13]:

$$\sigma_{ij} = \frac{1}{4\pi}[E_i E_j + H_i H_j - \frac{1}{2}(E^2 + H^2)\delta_{ij}], \tag{12}$$

indicating mutual repulsion between field-lines. This enables calculating the stress force in the general case and as an approximation above in Eq. (3). Our "stress conjecture" is therefore:

$$(F_s)_i = \frac{\partial \sigma_{ij}}{\partial x_j}, \tag{13}$$

again suggesting that the "ontological" viewpoint warrants consideration alongside the "instrumentalist" one. Moreover, we believe that, in the general case, Eq. (13) serves as a better estimate then (11) for the stress force.

Harpaz and Soker [10] straightforwardly relate the stress force's power to that of the emitted radiation:

$$P_s = -F_s a\Delta t = \frac{2q^2 a^2}{3c^3} = P_{rad}, \tag{14}$$

where $F_s$ is the stress force, and $P_s$ and $P_{rad}$ are the stress and radiation powers, respectively. In other words, the radiation power may be considered as originating from the stress force's work.

## 4.3. Deriving Electromagnetic Curvature from the Field's Mass and Stress

In fact, it is possible to derive a field's stress force from its mass and acceleration alone. Suppose that the field's mass in Eq.(10) is mechanically accelerated, in a direction deviating from the field-lines by $\theta$. The inertial "force" felt by the entire field would be

$$F_{inertial} = m_{field} a = \frac{q^2 a \sin\theta}{2c^2 r_q}. \qquad (15)$$

On the other hand, integrating the stress force in Eq.(11), we get

$$F_{stress} = 4\pi \int_{r_q}^{\infty} \frac{q^2 a \sin\theta}{4\pi c^2 r^2} dr = \frac{q^2 a \sin\theta}{c^2 r_q}, \qquad (16)$$

which differs from the force above by a factor of 2, originating from the crude averaging processes. The last equations suggest a very intuitive, "mechanical" view of electric interactions.

## 4.4. The Symmetry Conjecture in Terms of Field Charge and Mass

Our Symmetry Conjecture can now be rephrased as two modes of interplay between electricity and mass:
1. *Electromagnetic Charge-Mass Interplay:* When a charge accelerates, the acceleration is resisted not only by the mass within the charge itself (*e.g.*, an electron), but also, after some time, by the mass distributed over the surrounding field. Due to this temporal lag, the field curves, storing stress. The field-lines can release this stress either (*i*) by pulling the charge back to its original position, or (*ii*) by straightening back around the charge in its new frame. However, (*i*) is excluded by the relative smallness of the field's mass. Most of the stress, therefore, can be released only through (*ii*), namely, the lines release their stress through radiation as they straighten back. Viewing the field's configuration in terms of energy, the charge's kinetic energy first transforms into potential energy spread over the stressed field, and then to electromagnetic radiation, such that at any given moment the energy flow is

$$\Delta\varepsilon_k \Rightarrow \Delta\varepsilon_p + \Delta\varepsilon_{rad}. \qquad (17)$$

2. *Electrostatic Charge-Mass Interplay:* When two charges come close to one another, their field-lines bend each other, creating stress. This time, however, each field-line is connected not only to its own charge: Its remote end pushes/connects with the field-line of the other, equal/opposite charge. That other field-line, in turn, is connected to its own charge, *which has its own mass*. The two charges now constitute one system whose center of mass lies in the middle. It is now the fields, using each other as an Archimedes point, that have the upper hand in releasing the stress, resulting in the charges' acceleration towards/away from each other:

$$\Delta\varepsilon_p \Rightarrow \Delta\varepsilon_k + \Delta\varepsilon_{rad}. \qquad (18)$$

An equivalence, or even an approximate time-symmetry, thus relates (17) and (18), yet it is very crude. A more precise calculation would be needed in order to show that the energies involved in the two processes sum up the same way.

# 5. FAILURE OF A GENERAL PROOF: THE DIFFICULTIES AND THEIR LESSON

Obviously, our goal has not been fully attained. A rigorous proof for the equivalence of electromagnetic and electrostatic field-lines curvature should be of the following simple form:

1. Let $f(c_{mec})=a_{mec}$ be the function relating a charge's mechanical acceleration to the consequent curvature of its field-lines. Then obtain $a_{mec}$ from a given $c_{mec}$ for charge $q$.
2. Measuring $q$'s mass $m_q$, substitute $a_{mec}$ in (1) with $F/m_q$, then derive from $f(c_{mec})=F/m_q$ the force $F_{(mec)}$ that has produced $c_{mec}$.
3. Next, let the same curvature be produced electrostatically, by an external field induced by an appropriately configured charge, acting on our charge $q$. Produce, then, $c_{elec}= c_{mec}$.
4. Calculate the electric force that operates on $q$ to produce $c_{elec}$. Does $F_{elec}=F_{mec}$?
5. If (4) is the case, $c \Leftrightarrow a$ – QED.

This proof, we now realize, is impossible since according to Maxwell's equation, electrostatic fields have zero curl: $\nabla \times E = 0$. Therefore, (3) cannot be achieved, but we argue that even if we could perform (3), the expected result would not emerge in (4), neither the other way around.

But then, there are a few other differences between the electromagnetic and electrostatic curvatures which we have neglected so far, as well as some issues neglected in current electromagnetic theory. Let us review these issues before venturing to the desired proof.

## 5.1. Is the Curvature Equation Accurate?

First, a closer inspection of equation (3) suggests that it is only an approximation: It gives the field-lines curvature as independent of the distance from the charge, in obvious contrast with the physical mechanism underlying this curvature (see Sec.2) and even with the drawings supplied by the authors themselves (see Fig.2). The radius of curvature suggested by Harpaz is then accurate in the hyperbolic motion and also serves as a good estimation near the charge, yet it may not be adequate in long distances.

## 5.2. Energy Differences

Second, the electromagnetic and electrostatic curvatures differ in that the former is far weaker: Extreme and long-lasting acceleration is needed in order to produce a noticeable electromagnetic curvature, bearing in mind that the field-lines extend to huge distances, and moreover keep straightening back with the highest velocity, $c$. As pointed out in Sec.2, the electromagnetic curvature is a succession of momentary kinks that keep moving along the line outwards, ceaselessly dispersing energy in the form of radiation. One look at Larmor's equation (14) indicates the weakness of the electrodynamic power: it depends on: $c^{-3}$

The electrostatic curvature, in contrast, involves far smaller energies. Rather than numerous momentary kinks, the field-line presents one large, continuous and fixed curve. And rather than radiating energy, the field-line stores only the potential energy between the two charges as long as attraction/repulsion is prevented. Even when the charge is released, the ensuing acceleration is very small in comparison to that which gives rise to the electromagnetic curvature. The electromagnetic energy thereby released is similarly negligible.

No wonder, then, that the comparison between the electromagnetic and the electrostatic curvatures did not yield a simple equivalence.

## 6. THE SOURCE OF FAILURE: AN ASYMMETRY WITHIN THE SINGLE FIELD-LINE

Fortunately, all the above hints (and another one presented in Sec.10) converge into one hindrance to our Symmetry Conjecture: *A spatial asymmetry is inherent to the field-line itself*.

Recall, first, the conventional assumption (*i*) in Sec.1: Each field-line ends either at infinity or with an opposite charge somewhere. Consider, then, a charge $q$. In an empty universe, its field may go to infinity. But in a real universe full with matter, each of its field-lines must end in some remote opposite charge, such that they are straight and uniformly spread. We thus have numerous distant charges, $-q_1, -q_2, \ldots, -q_N$, each sharing a small fraction of the field-lines with $q$. These charges are henceforth dubbed "envelope charges."

The atomic nature of matter enables us to better identify these envelope charges. Although the Universe appears electrically neutral, this is so only at the macroscopic level. Microscopically, all matter is composed of positive and negative particles. At this level, therefore, matter is speckled with charges, most of them being atomic and molecular dipoles. These, when affected by a large electric field, become *polarized*: The dipole rotates with respect to the large charge, such that its opposite charge shares a line with it, while the dipole's other charge sends a free field-line to the opposite direction (Fig.4). The polarized dipole may further exert a similar effect on a nearby dipole and so on.

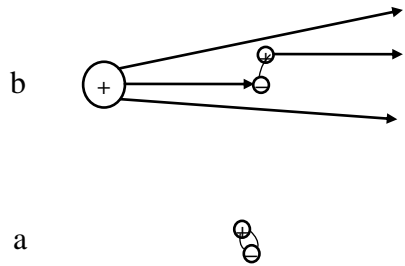

**FIGURE 4.** An electric dipole in a electrically neutral environment (a) and in the presence of a macroscopic charge (b).

This realistic refinement of the idealized textbook account will prove helpful in what follows.

Next consider the single field-line. To use Rowland's [8] term, the field-line is "anchored" to its charge. The nature of this anchorage is beyond the scope of this paper (and probably beyond the full understanding of present-day field theory). Yet one crude quantitative statement can be made about it: Nothing comparable to the force of this anchorage operates on the field-line's other end. Hence: *A field-line can be disconnected from the remote charge, but never be torn off its own charge*. This would contradict Gauss' law.

To realize that, consider again a charge which is merely *displaced* by a great distance. Where do its field-lines end now? Surely not with the charges in its previous location, otherwise, with large distances, they will have to be *curved*. A similar problem holds for an *inertially moving* charge.

Whatever the correct account of such cases, the asymmetry inherent to the single field-line is now obvious: When the charge is in motion, the field-line remains connected to its own charge, not to the remote one. The convention "an electric field-line starts with a positive charge and ends with a negative charge" is therefore flawed: The source and the distant charge are not on equal footing. This fact, namely, that the field-line is attached differently to the two charges, is the main hindrance to proving our Symmetry Conjecture: *The electromagnetic and electrostatic curvatures do not appear to be equivalent due to the asymmetry between the field-line's near and remote ends.*

In what follows we circumvent this hindrance somewhat nonchalantly, with the aid of a highly unrealistic *gedankenexperiment*. Next, we proceed to a more realistic demonstration of our Symmetry.

## 7. THE SYMMETRY PROVEN UNDER "SYMMETRIC ANCHORAGE"

Consider a charge $q$ at rest, all its field-lines extending to all directions with equal distances between them, each ending with a remote envelope charge somewhere. Next make the following unrealistic assumption: Each field-line is anchored to the envelope charge with the same force as to its own charge. Now let $q$ move, *slowly and inertially*, by a single nonelectric push applied directly on the charge.

Ignoring the initial acceleration, the charge's field-lines must begin curving from the moment that inertial motion has begun, and curve further as the motion proceeds, despite the fact that it is purely inertial. In other words, by virtue of our "symmetric anchorage" assumption, the field-lines can no longer be disconnected from the envelope charges. Now each of these envelope charges has its own mass. Therefore, together all the envelope charges, to which $q$'s field-lines are strongly anchored, offer considerable inertial resistance to $q$'s motion. Consequently, increasing stress occurs along $q$'s curving field-lines, increasingly resisting its motion. In other words, negative acceleration ensues till $q$ stops, and then is pulled back by almost the same acceleration, till the field-lines resume their initial straight form and $q$ returns almost to its initial location (the "almost"s refer to small motions that will nevertheless show up in the envelope charges, so some oscillations are expected). Throughout this process, curvature and acceleration appear and vanish together.

Is this curvature electromagnetic or electrostatic? Well, neither: It is an intermediate case. It is not electromagnetic because $q$'s initial motion is inertial. Neither is it electrostatic, because the envelope charges are too remote to affect $q$ electrostatically. It is by our imaginary assumption that the remote charges anchor the ends of $q$'s field-lines to themselves with the same force that anchors them to $q$. Under these circumstances, all $q$'s kinetic energy remains stored within its curved field-lines, which, unable to disconnect from the remote charges, cannot release this stress by straightening back and radiating energy. Once they pull $q$ back to its original position, *the Symmetry Conjecture strictly follows from energy conservation*: The stress force, given in Eq. (11) as acceleration's consequence, is now its cause:

$$F_s \equiv \frac{E^2}{4\pi R_c} = m_q a, \qquad (19)$$

in accordance with stage (5) of our "ideal proof" in Sec.5.

Indeed, a "symmetric anchorage" of the kind imagined here is often used as a way of understanding the strong force [15]. The quark and anti-quark are postulated to be connected by a "flux tube" – a configuration by which the field-lines connecting them are confined within a tube and do not spread out of the quarks. This makes the field density between the particles constant, while the force increases linearly with distance, by assumption (*v*) concerning the field-lines (Sec.1). In addition, the energy within the field grows quadratically with distance. The latter point explains why it is not possible to separate the quarks indefinitely: At some distance, the energy in the field will be large enough to produce a new quark and anti-quark pair between the original quarks. In other words, the field-lines mediating the strong force seem to be equally anchored on both quarks. The difference between the electromagnetic and strong forces can thus be understood as a consequence of symmetric *vs.* asymmetric anchoring.

Returning to the electromagnetic case, is a real experiment under "symmetric anchorage," possible? Perhaps a very unique configuration of charges may enable a realistic effect of this kind. In what follows, we proceed to the more modest goal of a partial proof for the Symmetry Conjecture.

## 8. THE ELECTROMAGNETIC CURVATURE TIME-REVERSED

Our previous setting was based on the unrealistic "symmetric anchorage" assumption. In reality, the envelope charges show no electrostatic interactions with their main charge. Only when the latter accelerates, it affects the envelope charges by its radiation. But then, even this interaction can be useful to our conjecture: If the *electromagnetic* interaction between a charge and its envelope charges is time-symmetric, perhaps a similar symmetry governs the more subtle, *electrostatic* connection between the charge and its envelope charges when they are at rest.

Consider then the following idealized case. Let the main charge $q$ be at rest in $\vec{r}(0) = (0,0,0)$, surrounded by $N$ equidistant envelope charges: $-q_1, -q_2, \ldots, -q_N$, also at rest. Next let $q$ accelerate, $\vec{a} = c\dot{\vec{\beta}}$ to velocity $\vec{v} = c\vec{\beta}$. Its field is given by:

$$\vec{E}(\vec{x},t) = q[\frac{(\hat{n}-\vec{\beta})(1-\beta^2)}{\kappa^3 R^2}]_{ret} + \frac{q}{c}[\frac{\hat{n}}{\kappa^3 R} \times \{(\hat{n}-\vec{\beta}) \times \dot{\vec{\beta}}\}]_{ret}, \quad (20)$$

where $\hat{n} = \frac{\vec{R}}{|R|} = \frac{\vec{x}-\vec{r}(t)}{|\vec{x}-\vec{r}(t)|}$, directed away from the charge's position $\vec{r}(t)$ to the observation point $\vec{x}$, and $\kappa = 1 - \hat{n} \cdot \vec{\beta}$. The "ret" subscript indicates that the bracketed quantity is to be evaluated at the retarded time $t' = t - R(t')/c$. We shall use the terms "velocity/acceleration field" for the field's first/second components.

Following $q$'s acceleration $\vec{a}$, each envelope charge $q_i$ undergoes a much smaller acceleration due to the acceleration field of $q$, which, in large enough distances, is greater than the velocity field:

$$\vec{a}_{qi} = \frac{q_i \vec{E}_{rad}}{m_i} = \frac{qq_i}{m_i c^2}[\frac{\hat{n}_i}{\kappa^3 R_i} \times (\hat{n}_i \times \vec{a})]_{ret}, \quad (21)$$

ignoring relativistic corrections. In other words, when $q$ accelerates, all the envelope charges undergo only minute accelerations, because the density of field lines in their vicinity is very small.

Our question now takes the following form: *Can we produce the reverse process, where the envelope charges accelerate first, giving rise to an "outside in" radiation, thereby imparting overall an equal acceleration on q?*

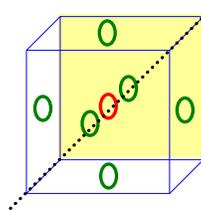

**FIGURE 5.** A charge surrounded by 6 equidistant envelope charges.

We first make two assumptions: (*i*) If the envelope charges are far enough, we can consider the simplified case where they are all equally remote from $q$, forming an imaginary spherical envelope. (*ii*) The envelope charges' direct electrostatic effect on $q$, namely, attraction, is negligible. Their "velocity field" depending on $1/R^2$ is much smaller than the "acceleration field" depending on $1/R$, hence can be ignored.

Consider, then, six envelope $-q/6$ charges surrounding $q$ at $\vec{r}_1(0)=(-d,0,0)$, $\vec{r}_2(0)=(d,0,0)$, $\vec{r}_3(0)=(0,-d,0)$, $\vec{r}_4(0)=(0,d,0)$, $\vec{r}_5(0)=(0,0,-d)$, $\vec{r}_6(0)=(0,0,d)$ as shown in Fig.4. Then accelerate these charges, $\vec{a}=c\dot{\beta}\hat{x}$ to $\vec{v}=c\beta\hat{x}$ for $t$ seconds to a distance $\ell$ from their original position. What would be their effect on the main charge $q$?

By Eq. (3),

$$\vec{E}_1(0,T) = \vec{E}_2(0,T) = 0$$

$$\vec{E}_3(0,T) + \vec{E}_4(0,T) = \vec{E}_5(0,T) + \vec{E}_6(0,T) = \frac{qad^2}{3c^2(\sqrt{\ell^2+d^2}-\beta\ell)^3}\hat{x}, \quad (22)$$

where $T = t + \frac{\sqrt{\ell^2+d^2}}{c}$.

The acceleration that $q$ acquires is

$$\tilde{\vec{a}} = \frac{q\sum_{i=1}^{6}\vec{E}_i}{m} = \frac{2q^2d^2}{3mc^2(\sqrt{\ell^2+d^2}-\beta\ell)^3}a\hat{x}. \quad (23)$$

The envelope charges' acceleration $a$ imparts on $q$ a proportional acceleration $\tilde{\vec{a}}$, where the constant of proportionality equals the ratio of its electrostatic and rest energies in the limit of $\ell \ll d$ which is relevant here,

$$\tilde{\vec{a}} = \frac{2q^2}{3mc^2d}a\hat{x}, \quad (24)$$

where the potential energy of the central charge appears explicitly.

Now, with larger numbers of envelope charges, we will get further contributions to $q$'s acceleration, eventually reaching the original acceleration $a$ when giving the above acceleration to all the envelope charges. The number of envelope charges should be:

$$N = \left\lceil \frac{N_G mc^2(\sqrt{\ell^2+d^2}-\beta\ell)^3}{q^2d^2} \right\rceil, \quad (25)$$

where $N_G$ is a geometrical factor. Then

$$\tilde{\vec{a}} = a\hat{x}, \quad (26)$$

which is just the mirror-equation of Eq. (21).

For the sake of completeness, let us consider also the case in which the number of envelope charges is infinite. We then replace the $N$ charges $-q_1$, $-q_2$, ... $-q_N$ with a continuous charge density per unit volume $\rho$, while maintaining the total charge $q = \int \rho dV$. In the limit of $\beta \ll 1$,

$$\tilde{\vec{a}} = \frac{8\pi q^2}{3mc^2d}a\hat{x}. \quad (27)$$

In both cases the symmetry is clear: (*i*) Acceleration of the main charge is followed by a radiation field, *i.e.*, field-lines curvature, which causes the envelope charges to accelerate as well. (*ii*) The entire process can occur *vice versa*.

## 9. THE PERSISTENT PROBLEM REVISITED: DOES A CHARGE RADIATE WHEN RESISTING GRAVITY?

Of course, no discussion of charge acceleration can avoid the notorious problem on which opinions are sharply divided to this day: If, by electromagnetism, an accelerating charge radiates, then, by the equivalence principle, it should radiate also when resisting gravity. Oddly, despite the question's simplicity, each of the two possible answers to it leads to serious difficulties: If the charge radiates, conservation laws are violated, whereas if it radiates only with respect to a free-falling observer, acceleration becomes relative. The debate therefore remains as intense as ever [3] [7] [8] [10]. It is only natural to expect our Symmetry Conjecture to have a bearing on this issue too.

## 10. THE CLUE: THE NON-RADIATING ACCELERATING PAIR

Our point of departure is a simple case in which the electromagnetic and electrostatic curvatures operate together, eventually cancelling one another in a way that illuminates their nature.

Let two charges $+q$ and $-q$ be separated by a distance that does not allow a noticeable electrostatic attraction between them. An important variable in this problem is the classical radii of the charges, here taken to be an electron and a positron, as it appears in the Thomson scattering formula

$$r = \frac{e^2}{mc^2}, \qquad (28)$$

which, as shown below, determine the transfer ratio of accelerations between the two particles.

Now let both charges accelerate at the same time and rate. Kislev and Vaidman [16] calculated their radiation:

$$E = \frac{qa}{c^2 l}, \qquad (29)$$

namely, each charge exerts some force that increases the other's acceleration:

$$F = \frac{q^2 a}{c^2 l} = \frac{r_e}{l} ma. \qquad (30)$$

Next, let the two particles be closer, enabling the electrostatic force to operate between them. They now form a dipole, with most of their field-lines strongly curved and *merged* within the distance separating them (see Fig.3). Only a small fraction of lines remains extending outside, hence the two charges' radiation is weaker the shorter the distance between them.

Finally, let the two particles be *maximally* close. They now form a *neutral* pair, which of course does not radiate.[1] Under such proximity, their combined inertial resistance is maximal: In order to accelerate one of them, double force is needed, so as to overcome also the mass of the other charge.

We now understand the failure in Sec.5: Only at the non-physical distance of $r_e$ between two charges (such that *all* the field-lines of one charge are connected to those of the other) can the charge accelerate the other charge's *entire* field. At a more realistic distance, a small portion of field-lines remains loose, hence the mutual acceleration is only partial.

This result adds further strength to our above conclusions concerning the acceleration of a *single* charge: Even in that case, the charge is interacting, though very weakly, with numerous opposite charges far away. The next question therefore follows: What happens if such a charge is accelerated simultaneously with its envelope charges?

---

[1]. In the case of a neutral atom, *e.g.*, the hydrogen atom, the uncertainty of the electron's position prevents the atom from being a dipole, and the electron totally screens the proton's electric field.

# 11. RADIATION OF A GRAVITY-RESISTING CHARGE

We now return to the infamous problem mentioned in section 9: Should a charge resisting gravity radiate? Our above analysis gives this question a fresh twist: *What role is played by the envelope charges that share this charge's field-lines?*

Merely raising the question seems to offer the needed hint: Very likely, most of these envelope charges also reside within the gravitational field that effects the main charge itself. Our question thus assumes a more general form: *How distant are the envelope charges from their main charge?*

We argue that the answer is: "*Not too far.*" The reason for this is evident from Sec.6 above: In an environment full with matter, the charge's field-lines do not end at infinity, neither with opposite charges faraway, but rather with nearby microscopic *dipoles*, from which new lines stretch outwards and so on.

This has an immediate bearing on our question: *If a charge accelerates together with its material environment, its field-lines curve only little, as their ends, attached to dipoles within that environment, accelerate with it.* Radiation is therefore barely noticed. A more precise account, in the spirit of [16], would invoke radiation emitted by both the main and envelope charges, but with destructive interference, leaving only little net radiation exchanged. It is only the accelerating frame's walls, whose dipoles are now polarized by the charge, that will radiate outwards.

In order to understand this argument let us consider three such co-accelerating reference-frames, each composed of a more exotic type of matter: (*i*) Conducting Matter: All the charge's field-lines end with nearby dipoles, while the same number of field-lines extend from the lab's external surface. Only the latter therefore radiate outwards. (*ii*) Plasma: Employing thermodynamic tools, Debye-Hückel [17] derived the Debye length used in plasma physics, and Fermi-Thomas derived the Fermi-Thomas screening wave vector used for dense electron gas [17]. These typical length scales emphasize the short range of electromagnetic fields in the "real world." Even for the intergalactic medium, the Debye length was found to be of order $10^5$ m. [18], again indicating the screening effect of envelope charges. (*iii*) Oppositely Charged Matter: If the main charge is symmetrically surrounded with small opposite charges that add up to the same magnitude, no field-lines extend outside, hence no radiation appears.

Back to ordinary matter, the conclusion is simple yet far-reaching: *For any charge, even macroscopic, within a dense material environment, the envelope charges are nearby dipoles to which its field-lines are attached.* This means that, *if a charge is used to test General Relativity within Einstein's proverbial elevator, most of its field-lines co-accelerate with it, hence no radiation is expected inside the elevator.* By the equivalence principle, the General-Relativistic conclusion follows: *Neither would there be radiation from a charge held fixed within a large gravitational field such as that of Earth.*

# 12. THE "THERMODYNAMIC RING" OF THE SYMMETRY CONJECTURE

While symmetries are positive indications for any physical model [9], the fact that physical reality is time-*a*symmetric requires our model to address this issue as well. Moreover, our above failure to produce a simple proof for the Symmetry Conjecture should turn out to be instructive for enabling our model to account for the real world's asymmetry.

Let us therefore return to the time-reversal of the electromagnetic curvature presented in section 8. Three consequences of that exercise are noteworthy, discussed below.

## 12.1. Entropy Increases in both the Normal and the Time-Reversed Process

The above time-reversal of radiation, must, of course, be highly unique: Extreme precision is required in order to make the field-lines begin to curve, from the envelope charges inwards, with perfect synchronization such that all the curvatures reach the charge simultaneously and bring about its acceleration. This, we now realize, resembles the famous, unphysical advanced solution to Maxwell's equations:

$$\vec{E} = \vec{E}(R+ct). \tag{31}$$

Why is such an advanced wave, converging on a charge prior to the latter's acceleration, never observed? Most authors argue that the phenomenon lies within the jurisdiction of the Second Law of Thermodynamics, and hence excluded as entropy-decreasing. In what follows we base this reasoning on the field-lines dynamics underlying the radiation.

Consider again, then, the time-reversed process invoked in Sec.8, where the envelope particles undergo minute accelerations first, bringing about a noticeable acceleration of the charge. While each envelope particles shares about one field-line with the main charge, it has several other field-lines, shared with other charges. Therefore, when this envelope charge accelerates, all these other field-lines also curve, but do not converge on main charge. Rather, they convey the radiation to the surrounding environment. Here again, entropy decreases only within the region populated by the main charge's field-lines, while increasing in all other directions (Fig. 6).

It seems, therefore, that the mere fact that a charge is connected to numerous other charges, suffices to set the stage for entropy increase in electromagnetic phenomena.

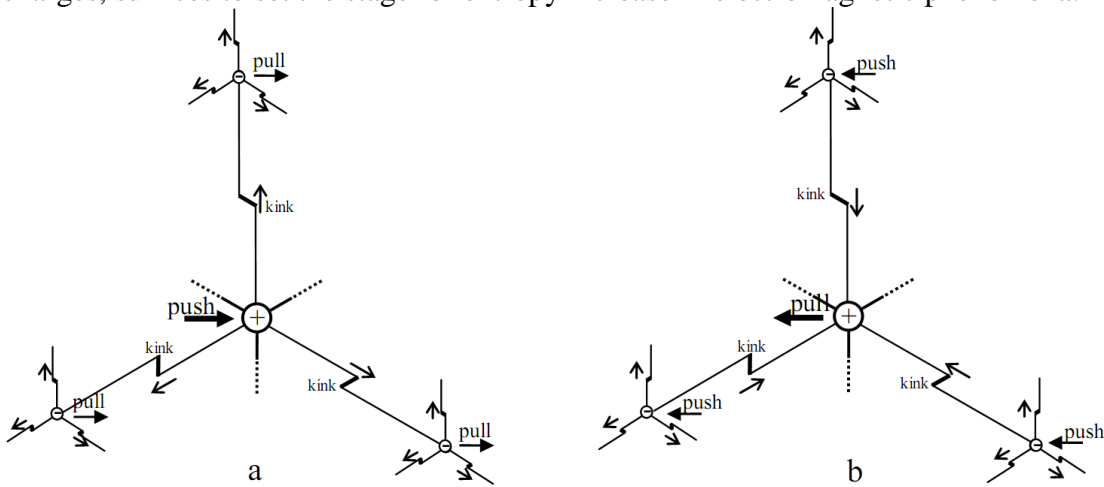

**FIGURE 6.** Normal (a) and time-reversed (b) electromagnetic radiation. Notice that, in the latter case, even though the small kinks on the envelope particles' field-lines are spatially inverted (left to right), they nonetheless spread *outwards*, as in the normal process.

## 12.2. A Local Realization of Maxwell's Equations' Advanced Solution

But, while an advanced wave like (31) is highly improbable, we now realize that something similar, though much smaller in scale, occurs with every electromagnetic absorption: *When a charge absorbs an electromagnetic wave from a distant charge, electromagnetic theory obliges the process to be, approximately, the mirror image of the radiation*. The electromagnetic wave, emitted by the first charge, is the familiar kink (Sec.2) reconnecting the old and new field-lines. Having spread over space, the spherical wave's front is now almost flat. Arriving in this form, it encounters several field-lines stretching out from the absorbing charge. Let us now follow the above intuitive reasoning to describe the next stages. The incoming, highly attenuated wave, slightly displaces the absorbing particle's field-lines, forming smaller kinks with their previous ends. These kinks now *converge* through the field-lines to their charge, which then accelerates (Fig.7).

So, while the emitted radiation is *dispersed*, the portion of it which is later absorbed by the other particle is *concentrated*. Of course the rest of the wave continues dispersing. Note that entropy increases even within absorption process: Once the absorbing charge accelerates, it emits back most of the radiation to the environment.

Finally, is it a mere coincidence that the conventional terms for a positive and negative charge are, respectively, "source" and "sink"? While their originators did not have thermodynamics in mind when conceiving them, these terms seem to reflect this subtle aspect of electromagnetism.

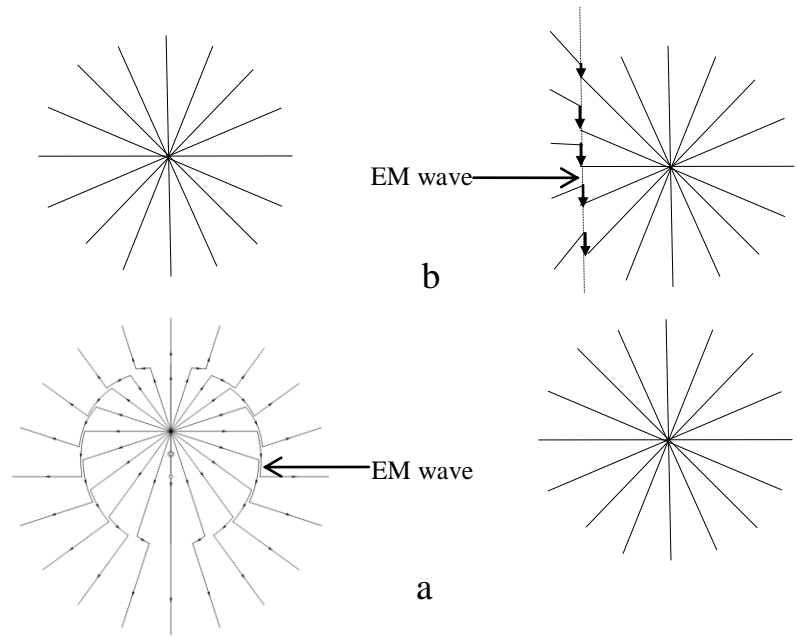

**FIGURE 7.** The symmetry between electromagnetic emission and absorption: (a) the conventional breaks and kinks occurring to the field-lines of the accelerated charge. (b) the opposite process occurring as the (now-plane) wave meets the field-lines of the absorbing charge, the kinks now converging rather than diverging.

## 12.3. The Inexistence of "Elastic Collision" as a Microscopic Origin of Time-Asymmetry

We believe that the above conclusions allow an even broader generalization, although very audacious: Even an interaction between two elementary particles has an irreversible ingredient. This is evident in the case of a particle and anti-particle pair, whose interaction ends in annihilation. But it holds also in the case of two identical charges that merely repel one another: Some energy is lost in the form of *radiation* emitted to the environment. This is the well-known Bremsstrahlung [19]. It is also, *inter alia*, a natural consequence of Rowland's [8] stressing that even a single particle is not a rigid body. Elastic collision, therefore, is a theoretical fiction, even at the *gedanken* level. We find it odd that the enormous literature on the origins of time-asymmetry [20] does not, to our knowledge, mention this straightforward origin.

Hence, the following basic principles of electromagnetism, namely,
   *i)* The field is enormously larger than its charge;
   *ii)* All field-lines end with remote opposite charges;
   *iii)* No influence propagates faster than light;
enable concluding that

*iv)* Even the most elementary interaction between two charges ends up with some of the kinetic energies transforming into radiation, irreversibly dispersed to the environment.

This account validates Einstein's position in the famous Einstein-Ritz debate [21] concerning the above absence of the advanced solution to Maxwell's equations. Whereas Ritz believed the reason to be some fundamental time-asymmetry inherent to electromagnetism, Einstein argued that the reason is, just like in macroscopic cases, purely probabilistic. Here, this probabilistic reasoning is shown to hold even when the source is an elementary particle.

## 13. COMPARISON WITH THE WHEELER-FEYNMAN ABSORBER THEORY

The Wheeler-Feynman "absorber theory" [22] [23] is celebrated for its bold resolution of the self-action problem, to which it responded by invoking explicit backwards-in-time interactions within the ordinary electromagnetic interaction. The emitter emits not one but two waves, one being the normal, retarded wave sent to the future while the other is an advanced wave going back to the past. The future absorber respond with another such pair of waves, the advanced one going back in time to the source. Then, a complex web of interference patterns between retarded and advanced actions removes all the unphysical parts, leaving the familiar, overall retarded wave.

As our model also seeks to reveal hidden symmetries beneath the electromagnetic process, a brief comparison between the two is in order.

Our "envelope charges" are identical to W-F's "absorbers." Indeed, W-F explicitly invoke one emitter and several absorbers, showing that, despite the latter's responding to the former's single wave with many advanced waves, the result is equivalent to one wave. On the other hand, our model does not invoke advanced action. We considered two time-reversed processes: The improbable converging macroscopic wave (Sec.12.1) and the normal wave-portion converging on the absorber (12.2). Both take place, however, in the positive time direction. Our model does not require advanced actions because the problems it addresses are more modest than those tackled by W-F. It does, on the other hand, describe the electromagnetic interaction in much greater detail, in that it studies the role played by the field-lines themselves. Should our reasoning prove to be sound, would it be worthwhile to reformulate the W-F model on this new basis?

## 14. EXPERIMENTAL CONSEQUENCES?

Returning to the question posed at the beginning of this paper, namely, Are the field-lines real physical objects? we should now consider a corollary: *Does it matter?* When two opposing interpretations give the same experimental predictions, the problem belongs to philosophy rather than to science. In what follows we show that the philosophical considerations that have guided us so far may lead to surprising empirical predictions.

First, let the two rival positions be fully phrased:

<u>The ontological interpretation:</u> Electric field-lines curve either when the charge accelerates or when a neighboring charge is present. Therefore, some common laws underlie the two curvatures.

<u>The instrumentalist interpretation:</u> Electric field-lines do not really curve in the presence of a neighboring charge. In reality, the interacting charges' field-lines remain straight, appearing to be curved only by the superposition principle.

Our first objection to the second alternative is, if we may, that it is ugly. It does not deny that field-lines curve under charge acceleration, because only one field is involved. Yet it does deny physical reality to the electrostatic curvature. This is highly unparsimonious. But apart from this aesthetic objection, could there be an experiment that distinguishes between the two options?

Let two equal and like charges move inertially towards one another until electrostatic repulsion makes them turn back. Once they come close enough, their field-lines gradually (appear to) curve as the two fields begin to overlap, and their motions undergo negative acceleration. More precisely, deceleration ensues till the charges halt, followed by their opposite acceleration until their original motions are reversed.

Following are the two competing accounts for this interaction (Fig.8).

<u>Ontological:</u> When the two fields come into contact, their field-lines begin to curve away from one another. This curvature produces pressure that proceeds from two field's overlap zone inwards each field, until reaching the charges, bringing about their negative acceleration. This declaration, however, does not always lead to further curvature. With the previously-curved lines, rather the contrary happens: As the charge's velocity is reversed, *the curved field-lines straighten up*, releasing the stress stored within them in the form of kinetic energy imparted back to the charges (8a).

<u>Instrumentalist:</u> When the two fields begin to overlap, the accompanying curvature is only *apparent*, an artifact of the superposition of the original vectors, which, in reality, *remain straight*. This pseudo-curvature therefore does not play any causal role in the interaction. Only when each field comes sufficiently close to the other charge, *directly* giving rise to the other charge's negative acceleration, do the latter's field-lines *really* curve. This curvature is electromagnetic, and, as pointed out in Sec.5.2, much weaker than the electrostatic one. But then, because the acceleration is negative, this *real* curvature is opposite to the pseudo-curvature: *The field-lines curve in the direction of the motion* (8a).

Is it possible to experimentally distinguish between these two accounts? Unfortunately, we are not able to present such a rigorous result yet. Although the instrumentalist account derived above is bizarre, counterintuitive and highly opposed to the ontological one (Fig.8), it is still possible that the superposition of two such oppositely-curved fields will give just the familiar Faraday pattern. Stricter mathematical analysis would be needed to find out whether this is indeed the case.

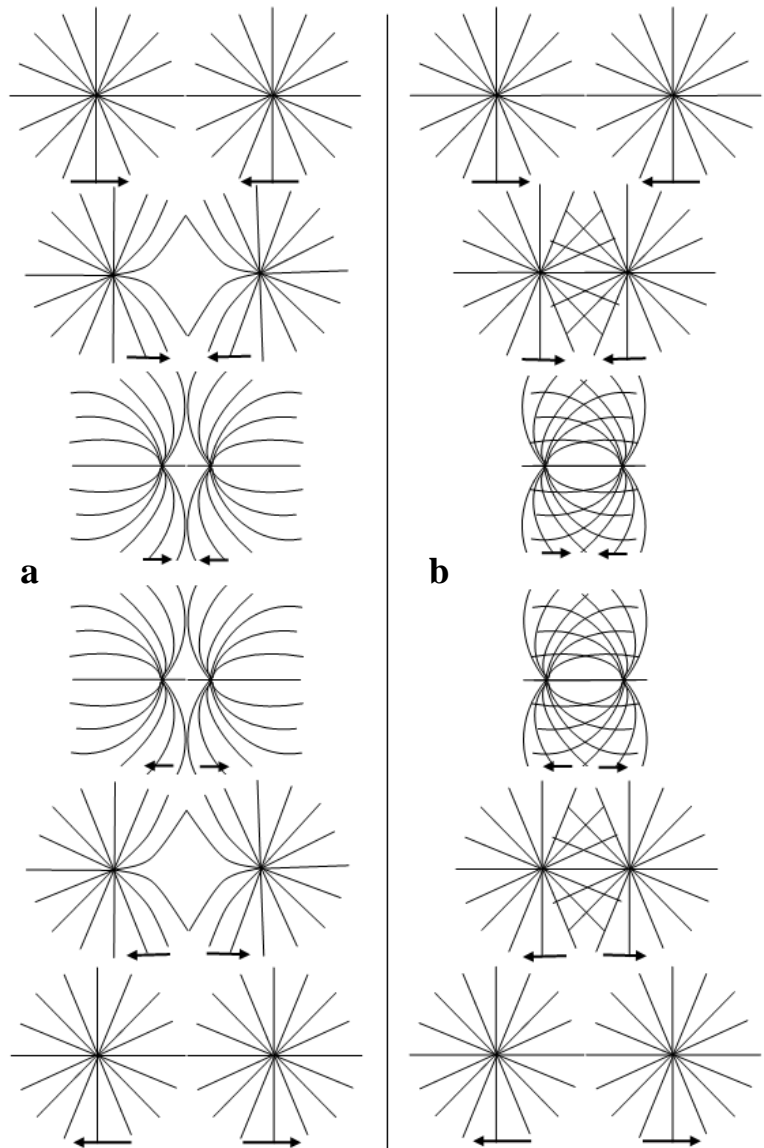

**FIGURE 8.** Two equal charges come close at inertial velocity and repel each other. a: The ontological account. The curving field-lines convey the pressure back to the charges. b: The instrumentalist account. The field-lines remain straight, the curvature being only an artifact of their superposition. Only upon declaration, real curvature appears, much weaker than the electrostatic one (but not shown in proper scale in this figure). This electromagnetic curvature is in the opposite direction, as if the two charges attract rather than repel one another.

Still, one essential difference between the two models seems to be bound to yield the hoped-for difference: If, according to the instrumentalist version, *the electrostatic field-lines curvature is not real*, then *it plays no causal role in the electric attraction/repulsion, appearing alongside it rather than preceding it*. The ontological model, in contrast, states that *the electrostatic curvature, indicating stress and mass-distribution changes within the field, is the very cause of the attraction/repulsion*.

This, we believe, should enable finding some cases where the two accounts give noticeable different predictions. We hope to present further advances on this issue in future works.

## 15. SUMMARY

The field-line is an essential aspect of the field concept itself, especially in electromagnetism. Several authors warn against taking this concept too literally, as if the field-lines are real physical objects. On the other hand, several other works indicate that the field has mass and is capable of being in stress. This strongly suggests that the field-line concept, when taken seriously, offers some insights into the very nature of the field phenomenon, which remains elusive even within present-day physics.

In this work we took the somewhat naïve view that field-lines are real physical objects. We find the conclusions derived from this approach, especially those concerning electromagnetism-gravity interactions, time-symmetry and irreversibility, encouraging enough to warrant further pursuing.

Should this reasoning prove sound, its extension to the quantum realm would be the next natural step. Moreover, the other physical fields may provide further aspirations, with gravity perhaps promising the most intriguing hints about the nature of spacetime.

## ACKNOWLEDGMENTS

It is a pleasure to thank Lawrence P. Horwitz, Joe Rosen, Alon Drori, Avi Gershon, Michael Bialy, Doron Grossman and Boaz Rubinstein for many illuminating discussions and helpful comments.